# Prediction of high-$T_c$ superconductivity in ternary lanthanum borohydrides

Xiaowei Liang[1], Aitor Bergara[2,3,4], Xudong Wei[1], Linyan Wang[1], Rongxin Sun[1], Hanyu Liu[5,6,*], Russell J. Hemley[7], Lin Wang[1], Guoying Gao[1,*], Yongjun Tian[1]

[1] *Center for High Pressure Science, State Key Laboratory of Metastable Materials Science and Technology, Yanshan University, Qinhuangdao 066004, China*
[2] *Departmento de Física, Universidad del País Vasco, UPV/EHU, 48080 Bilbao, Spain*
[3] *Donostia International Physics Center (DIPC), 20018 Donostia, Spain*
[4] *Centro de Física de Materiales CFM, Centro Mixto CSIC-UPV/EHU, 20018 Donostia, Spain*
[5] *International Center for Computational Method & Software and State Key Laboratory of Superhard Materials, College of Physics, Jilin University, Changchun 130012, China*
[6] *Key Laboratory of Physics and Technology for Advanced Batteries (Ministry of Education), International Center of Future Science, Jilin University, Changchun 130012, China*
[7] *Departments of Physics and Chemistry, University of Illinois at Chicago, Chicago IL 60607, USA*
* Corresponding author. E-mail address: gaoguoying@ysu.edu.cn;

hanyuliu@jlu.edu.cn

**ABSTRACT**

The study of superconductivity in compressed hydrides is of great interest due to measurements of high critical temperatures ($T_c$) in the vicinity of room temperature, beginning with the observations of $LaH_{10}$ at 170-190 GPa. However, the pressures required for synthesis of these high $T_c$ superconducting hydrides currently remain extremely high. Here we show the investigation of crystal structures and superconductivity in the La-B-H system under pressure with particle-swarm intelligence structure-searches methods in combination with first-principles calculations. Structures with six stoichiometries, LaBH, $LaBH_3$, $LaBH_4$, $LaBH_6$, $LaBH_7$ and $LaBH_8$, were predicted to become stable under pressure. Remarkably, the hydrogen atoms in $LaBH_8$ were found to bond with B atoms in a manner that is similar to that in $H_3S$. Lattice dynamics calculations indicate that $LaBH_7$ and $LaBH_8$ become dynamically stable at pressures as low as 109.2 and 48.3 GPa, respectively. Moreover, the two phases were predicted to be superconducting with a critical temperature ($T_c$) of 93 K and 156 K at 110 GPa and 55 GPa, respectively ($\mu^*$=0.1). Our results provide guidance for future experiments targeting new hydride superconductors with both low synthesis pressures and high $T_c$.

**Introduction**

Exploration of superconductivity in materials at ever increasing temperatures is a burgeoning research topic in condensed matter physics, chemistry, and materials science. Conventional electron-phonon coupling considerations point to compressed hydrogen-rich materials as excellent candidates for superconductors having high critical temperatures ($T_c$'s) due to the potential for formation of atomic hydrogen lattices in which the low mass leads to both high vibrational frequencies and strong electron-phonon coupling. As originally proposed by Ashcroft[1], this concept has inspired numerous studies (see Refs.[2-6] for reviews), specifically the recent progress on pressurized hydrides predicted and observed to have high $T_c$'s above 200 K in pursuit of superconductivity at, or even above, room temperature[7-21]. However, pressures in the megabar range (>100 GPa) are required to synthesize and stabilize the high-$T_c$ hydrides considered so far. For example, high-$T_c$ superconductivity was established above 170 and 166 GPa for clathrate metal hydrides $LaH_{10}$[16] and $YH_6$[19], and near 155 and 267 GPa for $p$-block element hydrides $H_3S$[13] and C-S-H[21], respectively, where the pressures are those of the reported $T_c$ maxima. Given the very high pressures required to create these high critical temperatures, the pursuit of high-$T_c$ superconductivity in hydrides that can persist in stable or metastable compounds at lower, and even ambient, pressure remains an important goal.

The stability of binary hydrides having potential superconducting $T_c$'s above 100 K has been largely limited to pressures above 100 GPa[22]. For example, synthesis of superhydride $UH_7$ has been reported at a low pressure of 31 GPa, but its $T_c$ is estimated to be 44 K[23,24]. The lowest pressures reported for stabilization of a superhydride include those of $CeH_9$ at 80 GPa[25] and $BaH_{12}$ at 75 GPa[26] for atomic and molecular-based hydrogen structures, respectively. Predictions of lower pressure stability of hydrogen-rich binary hydrides include that of $RbH_{12}$, which is calculated to be stable at 50 GPa with a $T_c$ near 115 K[22]. With the additional degrees of freedom made possible by expanding the chemical space available, ternary hydrides are receiving growing interest as means both to increase $T_c$ and to enhance stability over a broader range of pressures. As such, a $T_c$ of 287 K has been reported in a C-S-H mixture at about 267 GPa, while the structure and composition for the high-$T_c$ phase remain unclear[21]. Theoretical calculations predict that hydride perovskite structures based on the above elements could be a route to stabilizing lower pressure hydride superconductors, for example, by sublattice replacement of $SH_3$ with $CH_4$ in $H_3S$ to produce structures of composition $CSH_7$ with predicted dynamical stability, and therefore kinetic stability, at lower pressures than that of pure $H_3S$[27,28]. Lower level $CH_4$ substitution in the material, either as stoichiometric compounds or doped structures, could enhance low-pressure stability, as well as significantly enhance $T_c$ as recently predicted for the C-S-H superconductor[29]. These results further suggest that ternary hydride systems may be a useful venue for discovering high-$T_c$ superconductors at low pressures.

Theoretical studies indicate that adding Li and Ca to the B-H binary stabilizes phases that accommodate more H atoms, leading to ternary hydrides with higher $T_c$'s relative to those found for B-H phases[30-34]. Given its larger ionic radius, La can accommodate more atomic H (*e.g.,* as a superhydride) compared to Li and Ca[11]. Therefore, the stable phases with higher H content might be obtained in the La-B-H system under pressure. Moreover, a recent experimental study reported evidence for superconductivity at and above room temperature in La-hydride samples upon thermal annealing[35]. Given that ammonia borane ($NH_3BH_3$) was used as a hydrogen source, the formation of La-H phases containing boron was suggested as giving rise to the high $T_c$[16, 35].

In this study, we examine theoretically high-pressure structures, stability, and superconducting properties of stoichiometric La-B-H phases, with a focus on lower pressure stability. Detailed study of phases with composition $LaBH_x$ (x=1-10) reveals intriguing H-rich $LaBH_7$ ($P\bar{3}m1$) and $LaBH_8$ ($Fm\bar{3}m$) structures containing $BH_6$ and $BH_8$ units, respectively. Moreover, $LaBH_7$ and $LaBH_8$ are dynamically stable at pressures as low as 109.2 and 48.3 GPa, with predicted $T_c$'s of 93 and 156 K at 110 and 55 GPa, respectively. Our results indicate that continued exploration of ternary hydrides in these and related chemical systems may be an effective route to realizing a high-temperature superconductivity at lower, or even ambient, pressure.

**Results and Discussion**

Before investigating the phase stability of ternary La-B-H compounds under pressure, we first assessed information about the La-H, B-H and La-B binaries. The high-pressure behavior of the B-H[32-34] and La-H[11,12,36-38] binaries has been well-studied in recent years, whereas information on the La-B system under pressure is lacking. We therefore performed structure-search calculations for $La_nB_m$ (n=1, m=1-6; n=2, m=1) with system sizes containing up to 4 or 8 formula units (f.u.) per simulation cell at pressures of 100-300 GPa. To identify the stability of different stoichiometries, convex hulls were constructed by calculating the formation enthalpies for predicted $La_nB_m$ structures relative to the elemental La and B (Fig. S1). All the stoichiometries were found to have negative formation enthalpies within 100-300 GPa, showing that they are thermodynamically stable with respect to decomposition into La and B elements. The increase in formation enthalpy indicates that the stability of the phase decreases with increasing pressure; in addition, a stoichiometry located on the hull is stable with respect to other binary compounds, otherwise it is metastable. From the convex hull calculations, we found that LaB, $LaB_4$, $LaB_5$ and $LaB_6$ are stable at 100 GPa. At 200 GPa, LaB and $LaB_5$ remain on the convex hull, whereas $LaB_4$ and $LaB_6$ were predicted to possibly decompose into $LaB+LaB_5$ and $LaB_5+B$, respectively. At 300 GPa, the stable stoichiometries are LaB, $LaB_3$ and $LaB_6$. Notably, LaB was found to remain thermodynamically stable throughout the entire pressure range studied.

Calculated La-B-H ternary phase diagrams at 100-300 GPa are presented in Fig. 1. All the ternary hydrides are stable relative to dissociation into elements at the pressures studied. Moreover, all the ternary hydrides studied have lower enthalpies than those of LaB and $H_2$, suggesting that they are stable against decomposition into LaB and $H_2$ (Fig. S2). We note that LaBH, $LaBH_4$ and $LaBH_6$ fall on the 3D convex hull at 100 GPa, indicating that they are also stable phases with respect to decomposition into binary and other ternary phases. At 150 GPa, an additional composition, LaBH$_7$, appears on the convex hull. With further increase in pressure to 200 GPa, the originally stable LaBH$_6$ is predicted to decompose into LaBH$_4$ and LaBH$_7$, and the higher H content LaBH$_8$ begins to become stable. At 300 GPa, the formation enthalpy of LaBH$_8$ is increasingly negative. LaBH and LaBH$_7$ become metastable phases with higher enthalpies relative to 1/4 LaB$_3$+1/4 LaB+1/2 LaH$_2$ and 1/4 LaBH$_4$+3/4 LaBH$_8$, respectively, and LaBH$_3$ falls on the convex hull. To determine accurate stability pressures, we also plot the specific enthalpy curves of LaBH$_4$, LaBH$_6$, LaBH$_7$ and LaBH$_8$ relative to other compounds (Fig. S3). After including zero-point energy corrections (Fig. S4)[39], we found that LaBH$_4$ remains thermodynamically stable in a $P2_1/m$ structure within the entire pressure range studied. $C2/c$-LaBH$_6$ is stable below 134 GPa. For LaBH$_7$, the $P\bar{3}m1$ structure is predicted to become stable against dissociation into other stoichiometries at pressures of 103-233 GPa. LaBH$_8$ is predicted to crystallize in the cubic $Fm\bar{3}m$ structure, which is stable relative to LaBH$_7$ and H$_2$ above 161 GPa.

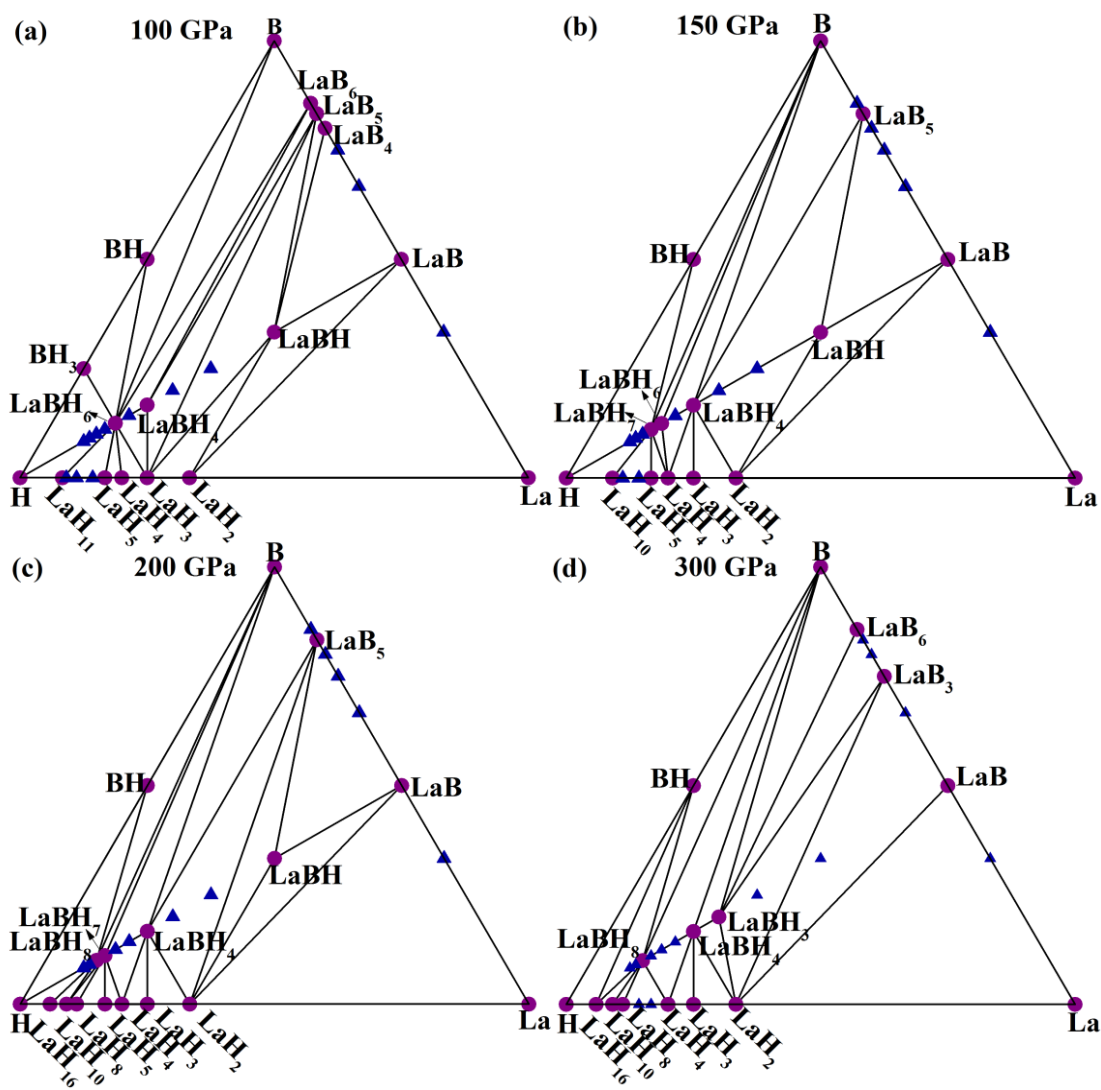

**Fig. 1** Calculated convex hull of the La-B-H system at 100, 150, 200 and 300 GPa.

Thermodynamically stable and metastable stoichiometries are shown as purple circles and blue triangles, respectively.

The predicted stable structures of the La-B-H system are shown in Fig. 2. LaBH adopts the hexagonal *P*6/*mmm* structure, in which B atoms form honeycomb sheets and B, La, and H atomic layers are alternately arranged. In $LaBH_3$ and $LaBH_4$, zigzag B chains stretch along specific directions while B atoms are surrounded by H atoms to form covalent bonds. Since more H atoms are filled in *C*2/*c*-$LaBH_6$, no bonds exist between B atoms and each B atom forms a $BH_4$ unit with the adjacent four H atoms. With the increasing H content of $LaBH_7$, each B atom accommodates six H atom to form $BH_6$ units. The $BH_6$ groups are located on the vertices and edges of the hexagonal structure, with the $BH_6$ units distributed on the edges connected to each other by H atoms. The H-richer $LaBH_8$ assumes a high-symmetry $Fm\bar{3}m$ structure in which B atoms accommodate all the H atoms to form $BH_8$ covalent units that occupy the octahedral interstices of the face-centered cubic lattice formed by La atoms. In $H_3S$, the S atoms located on a body-centered cubic lattice with each S atom covalently bonded to the surrounding six H atoms. The H atoms in $LaBH_8$ are found to bond with B in a manner that is similar to that of S in $H_3S$. In addition, the atomic positions of La and eight H atoms in $LaBH_8$ are the same as those of La and eight of the H atoms in $LaH_{10}$. Given the high $T_c$ of $H_3S$ and $LaH_{10}$, this similarity in bonding and high symmetry structure suggests interesting superconducting properties of $LaBH_8$ as well.

Lattice dynamics calculations were carried out for the phases in the pressure ranges of their predicted thermodynamic stability. The lack of imaginary frequencies in the calculated phonon dispersion curves indicates that all structures are dynamically stable within the harmonic approximation (Figs. 5 and 5S). On the other hand, phonon softening is evident for $LaBH_7$ and $LaBH_8$, an effect that can enhance the EPC[11]. With decreasing pressure, these phonon modes further soften and eventually have imaginary frequencies. Figure S6 shows the frequency of the softest mode as a function of pressure. In contrast to $LaH_{10}$, $LaBH_7$ and $LaBH_8$ maintain dynamical stability to pressures as low as 109.2 and 48.3 GPa, respectively, and the latter is much lower than that predicted for other H-rich hydride superconductors. By analyzing the eigenvectors of soft modes for $LaBH_7$ at the *M* point and $LaBH_8$ at the *Γ* point (Fig. S7), it is found that the structural instability mainly stems from the vibration of H atoms. Similar behavior was found for the clathrate superhydride $LaH_{10}$[11,36].

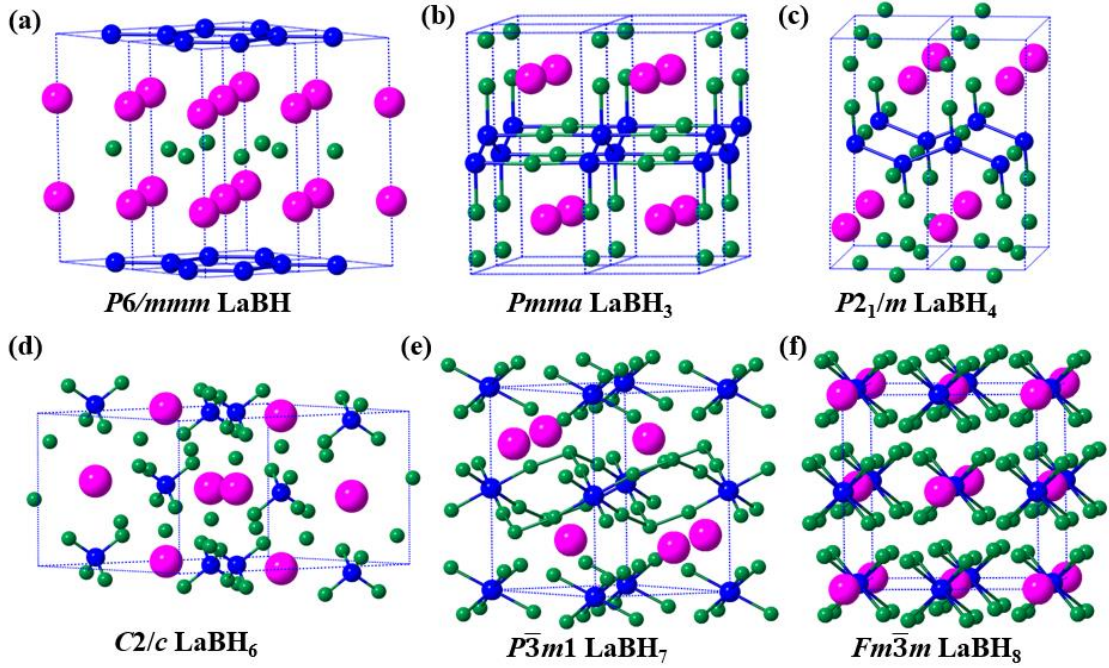

**Fig. 2** The predicted crystal structures of ternary hydrides under pressure. (a) $P6/mmm$-LaBH, (b) $Pmma$-LaBH$_3$, (c) $P2_1/m$-LaBH$_4$, (d) $C2/c$-LaBH$_6$, (e) $P\bar{3}m1$-LaBH$_7$ and (f) $Fm\bar{3}m$-LaBH$_8$. Magenta, blue and green balls represent La, B and H atom, respectively.

To understand the origin of relatively low-pressure stability of $P\bar{3}m1$-LaBH$_7$ and $Fm\bar{3}m$-LaBH$_8$, we explored the bonding of these structures by calculating the electron localization function (ELF)[40,41] and Bader charge transfer[42] among atoms. ELFs with an isosurface of 0.6 are shown in Figs 3a and b for the two phases at 110 and 55 GPa, respectively. Electron density at the La atoms is due to their inner valence shells. Many electrons are clearly localized between B and H atoms and closer to the H atoms. The ELF slice in the (110) plane containing La, B and H atoms for LaBH$_7$ and LaBH$_8$ (Fig. 3c) also shows that the ELF values between B and H atoms gradually increase toward H atoms, suggesting the polar covalent character of the B-H bond. In LaBH$_7$, atom H2 appears to form a covalent bond to H1 with an ELF value of 0.64 connecting BH$_6$ units on the edges. In both phases, the ELF values at the center of the shortest La-H and La-B are below 0.3, indicative of an ionic character between La and B-H units.

Bader charge calculations show that electrons transfer from La and B to H atoms. In LaBH$_7$, each La atom and B atom located on the vertex and edge of the lattice loses 1.44, 1.37 and 1.16 electrons, respectively. Correspondingly, each H atom in BH$_6$ at the vertex (H3) and edge (H1) accepts 0.45 and 0.39 electrons, respectively. The H atom (H2) that only bonds with a H atom gets 0.15 electrons. The existence of the H1-H2 covalent bond weakens the B-H1 bond connected to it. In LaBH$_8$, each La and B atom transfers 1.47 and 1.04 electrons to eight H atoms, respectively. With increasing pressure, the number of electrons transferred by La and B atom decreases and increases, respectively. In H-rich $Fm\bar{3}m$-LaH$_{10}$, the H atoms are linked by weak covalent bonds to form cage structures. As pressure decreases, the H-H interactions in cages become weaker, which leads to an instability of the H cages[11,36]. The results suggest that in these H-rich La-B-H compounds, strong interactions between B and H atoms, and between La and B-H units, play an important role in their relatively low-pressure dynamical stability.

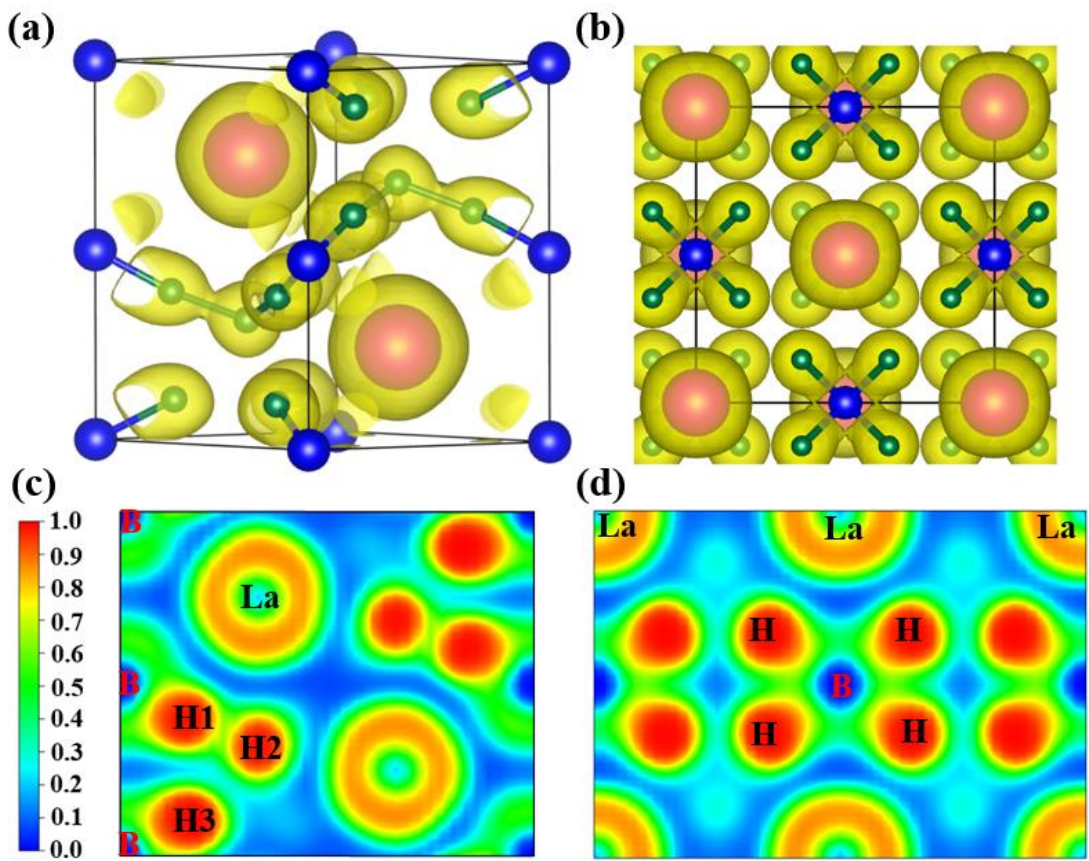

**Fig. 3** The Calculated ELF with isosurface value of 0.6 and ELF in the (1 1 0) plane for (a) $P\bar{3}m1$-LaBH$_7$ and (b) $Fm\bar{3}m$-LaBH$_8$ at 110 and 55 GPa, respectively.

We further investigated the electronic properties of the stable structures found in the La-B-H system. The calculated electronic density of states (DOS) for $P6/mmm$-LaBH, $Pmma$-LaBH$_3$, $P2_1/m$-LaBH$_4$ and $C2/c$-LaBH$_6$ within their ranges of pressure stability are shown in Fig. S8. The electronic DOS at the Fermi level indicates that they are all metallic. $P6/mmm$-LaBH, $Pmma$-LaBH$_3$ and $P2_1/m$-LaBH$_4$ all have relatively high DOS values at the Fermi level. However, this metallicity is mainly derived from the contribution of La and B atoms: there is a negligible H contribution to the DOS at Fermi level, which is unfavorable to superconductivity. In $C2/c$-LaBH$_6$, the Fermi level falls at the valley of the electronic DOS, showing poor metallicity.

We further focused our investigation on the H-richer LaBH$_7$ and LaBH$_8$ compounds. Figure 4 illustrates the calculated electronic band structures and DOS of $P\bar{3}m1$-LaBH$_7$ and $Fm\bar{3}m$-LaBH$_8$ at 110 and 55 GPa, respectively. They are the metallic phases with some bands crossing the Fermi level. In LaBH$_7$, a flat band with more localized electronic states appears near the Fermi level at the $\Gamma$ point, which might enhance the electron-phonon interactions. "Flat-steep" band features are beneficial for superconductivity[43]. As such, the steep and flat bands are found for LaBH$_8$ along the $\Gamma$-X and X-W directions near the Fermi level, respectively. The trend of the band dispersions for LaBH$_8$ in 200 GPa is similar to that in 55 GPa (Fig. S9). In addition, the contribution of H atoms to the DOS at the Fermi level exceeds that of La and B atoms in LaBH$_7$, and the metallicity are dominated by H atoms in LaBH$_8$, which suggest that $P\bar{3}m1$-LaBH$_7$ and $Fm\bar{3}m$-LaBH$_8$ may be high-$T_c$ superconductors.

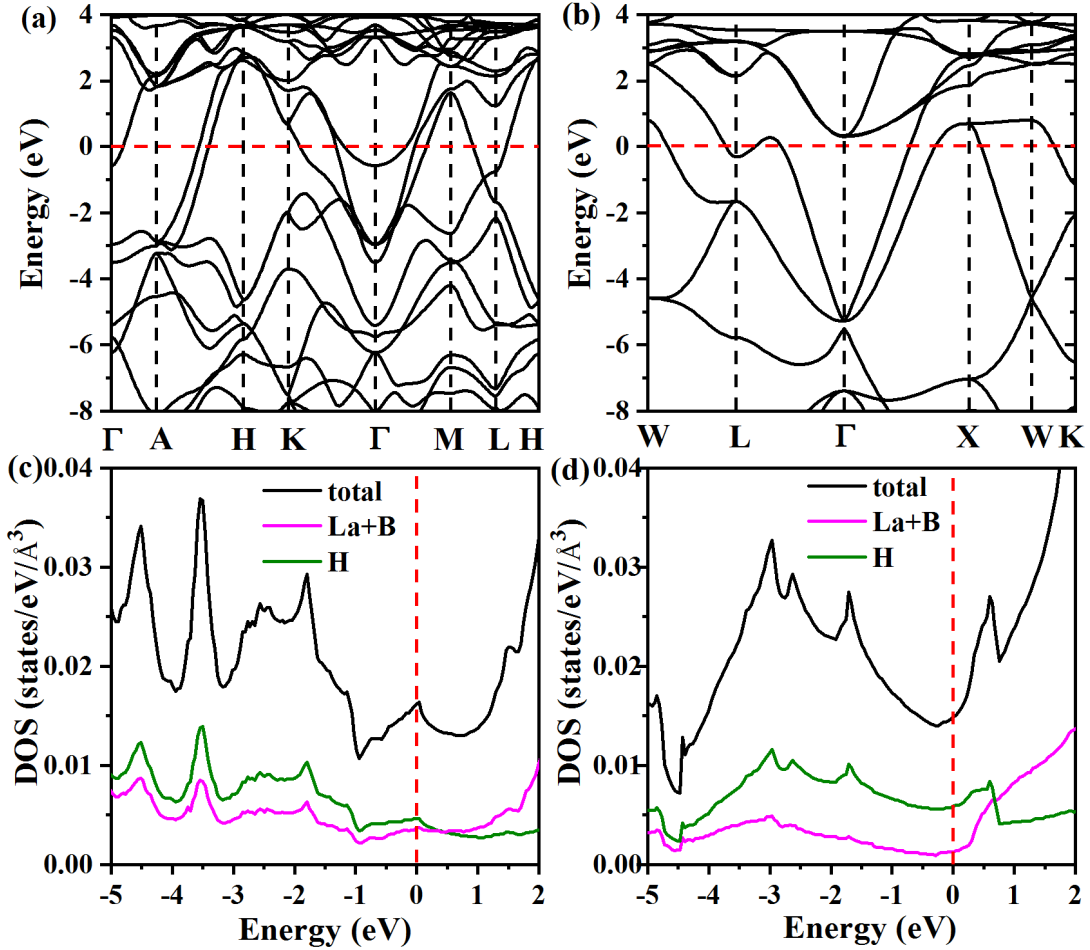

**Fig. 4** The calculated electronic band structure and density of states of $P\bar{3}m1$-LaBH$_7$ and $Fm\bar{3}m$-LaBH$_8$ at 110 and 55 GPa, respectively.

Given their promising electronic properties, we calculated the superconducting properties of LaBH$_7$ and LaBH$_8$. We calculated their phonon spectra, projected phonon DOS, Eliashberg phonon spectral function $\alpha^2F(\omega)/\omega$ and integral $\lambda(\omega)$ for the two phases at 110 and 200 GPa, respectively (Fig. 5a, b). Similar to the hydrides studied previously, the projected phonon DOS can be separated into three regions. The La atom with the heaviest atomic mass dominates the low-frequency region, whereas the vibrations of the B and H atoms are associated with the mid- and high-frequency phonon branches, respectively. The spectral function $\alpha^2F(\omega)/\omega$ for LaBH$_7$ is mainly distributed below 30 THz, especially between 8-15 THz (Fig. 5a), which results in an EPC constant $\lambda$ of 1.46 at 110 GPa. However, the value of the phonon DOS between 8-15 THz is negligible. Further analysis reveals a soft mode in this frequency range with a potentially large EPC contribution. The distribution of the EPC strength on the different phonon modes are also plotted with the spectra. The soft mode associated to H atoms below 20 THz around the *M* point shows a quite large EPC. Similarly, for LaBH$_8$ the calculated EPC $\lambda$ is 0.72 at 200 GPa, and the contribution to $\lambda$ of the vibrations related to H atoms above 30 THz accounts for 83% of the total value. The soft mode near 30 THz at $\Gamma$ makes an important contribution to the EPC. Previous studies of related superconducting hydrides indicate that the total EPC may be enhanced by further phonon softening induced by decompression toward the structural instability predicted by this harmonic approximation of the lattice dynamics[7,11]. Calculations for LaBH$_8$ indicate that $\lambda$ increases to 1.97 and 2.29 near its predicted instability at 55 (Fig. 5c) and 50 GPa (Fig. S10), which are comparable with the value of 2.19 found for H$_3$S at 200 GPa. The main contribution to the strong EPC at 55 GPa arises from soft modes below 20 THz associated with H atoms (Fig. 5c).

We adopted the Allen-Dynes modified McMillan equation to estimate the $T_c$ of $P\bar{3}m1$-LaBH$_7$ and $Fm\bar{3}m$-LaBH$_8$ at different pressures (Table I)[44]. For LaBH$_7$, the calculated $\lambda$ and phonon frequency logarithmic average $\omega_{\log}$ is 1.46 and 837 K at 110 GPa, leading to a $T_c$ of 93 K with $\mu^*$=0.1. As pressure decreases from 200 to 100, 55 and 50 GPa, the calculated $\lambda$ for LaBH$_8$ increases from 0.72 to 1.11, 1.97 and 2.29, whereas $\omega_{\log}$ decreases from 1557 to 1189, 807 and 692 K. As a result of these two effects, the calculated $T_c$ first increases from 58 to 115 K and then decreases to 108 K assuming $\mu^*$=0.1, which follows the trend of $\lambda$ and $\omega_{\log}$ with pressure, respectively. Since the $\lambda$ of LaBH$_8$ at 55 and 50 GPa are much greater than 1.5, the $T_c$ values were also calculated by numerically solving the Eliashberg equation[45], which gives $T_c$ values of 156 and 154 K with $\mu^*$=0.1, respectively.

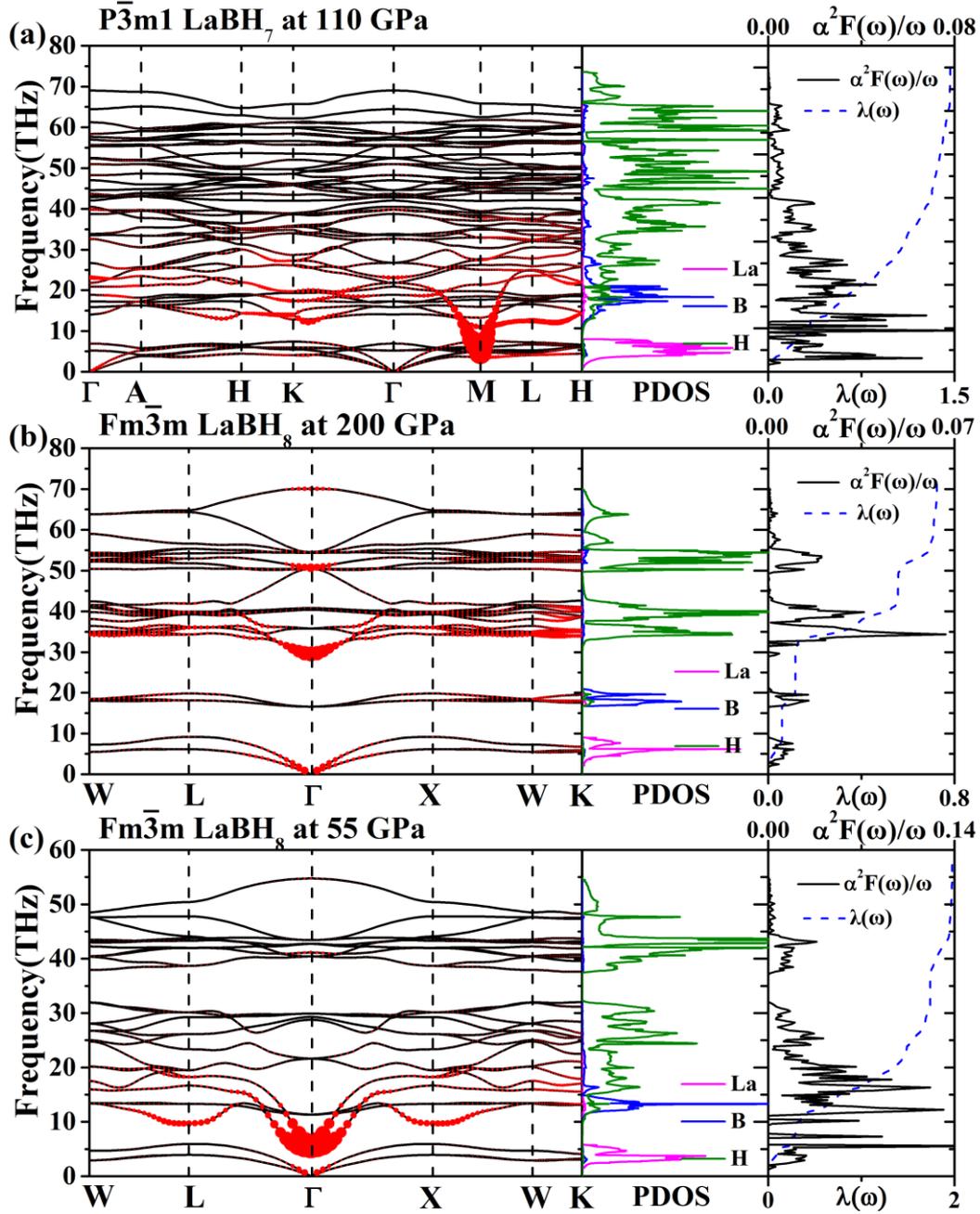

**Fig. 5** Calculated phonon dispersion curves (red circle area proportional to associated EPC), projected phonon density of states (PDOS), the Eliashberg phonon spectral function $\alpha^2F(\omega)/\omega$ and its integral $\lambda(\omega)$ of (a) $P\bar{3}m1$-LaBH$_7$ at 110 GPa, (b) $Fm\bar{3}m$ LaBH$_8$ at 200 GPa and (c) $Fm\bar{3}m$-LaBH$_8$ at 55 GPa.

**Table I.** The calculated electron-phonon coupling parameter $\lambda$, phonon frequency logarithmic average $\omega_{\log}$ and critical temperature $T_c$ ($\mu^*$=0.1-0.13) from Allen-Dynes modified McMillan and Eliashberg equations for $P\bar{3}m1$-LaBH$_7$ and $Fm\bar{3}m$-LaBH$_8$.

| Phase | Pressure (GPa) | $\lambda$ | $\omega_{\log}$ (K) | $T_c$ (K) $\mu^*$=0.1-0.13 McMillan | $T_c$ (K) $\mu^*$=0.1 Eliashberg |
|---|---|---|---|---|---|
| LaBH$_7$ ($P\bar{3}m1$) | 110 | 1.46 | 837 | 93-85 | — |
| LaBH$_8$ ($Fm\bar{3}m$) | 200 | 0.72 | 1557 | 58-45 | — |
|  | 100 | 1.11 | 1189 | 96-84 | — |
|  | 55 | 1.97 | 807 | 115-107 | 156 |
|  | 50 | 2.29 | 692 | 108-102 | 154 |

**Conclusions**

Density functional theory-based structure-search calculations have identified six phases in the ternary La−B−H system at pressures of 100-300 GPa that are potential targets for experimental synthesis. Most significant are the predictions of stability of H-rich $P\bar{3}m1$-LaBH$_7$ at 103-223 GPa and $Fm\bar{3}m$-LaBH$_8$ above 161 GPa, with the latter calculated to be dynamically stable as low as 48.3 GPa. Structural trends among these phases are observed as the H content increases. In LaBH, the B atoms form graphene-like layers, whereas in LaBH$_3$ and LaBH$_4$, the B atoms not only bond with each other to form zigzag chains, but bond with H atoms. In LaBH$_6$ and LaBH$_7$, there are no B-B bonds and B atoms are coordinated by Hs to form BH$_4$ and BH$_6$ units. LaBH$_8$ is stable in the high-symmetry $Fm\bar{3}m$ structure, in which the B atoms accommodate all the H atoms to form BH$_8$ units. The La atom acts as an electron donor in the structures to stabilize the higher H content B-H units. Moreover, EPC calculations show that LaBH$_7$ and LaBH$_8$ are potential superconductors. Softening of phonons dominated by H-atom vibrations in these structures makes a large contribution to superconductivity. The estimated $T_c$ of LaBH$_7$ is 93 K at 110 GPa, whereas the $T_c$ of LaBH$_8$ is calculated to be as high as 156 K at 55 GPa. The expanded range of dynamical stability to low pressures together with its predicted relatively high $T_c$ make $Fm\bar{3}m$-LaBH$_8$ a promising candidate superconductor for low-pressure stabilization experiments. A similar result for LaBH$_8$ is reported in a paper that appeared after the present calculations were completed[46]. Additional chemical substitution of these phases could be used to enhance both $T_c$ (*e.g.,* by electron or hole doping) or structural stability at still lower pressures. Additional theoretical work could explore potential anharmonic and quantum effects on the stability and the calculated critical temperatures[36,37,47] The present study is thus expected to stimulate further research on ternary and more complex superconducting hydrides with high critical temperatures and expanded ranges of stability.

**Computational details**

The structure searches of LaB and LaBH$_x$ (x=1-10) with simulation cells containing up to 4 formula units (f.u.) were performed at 100, 200 and 300 GPa using the particle swarm optimization technique implemented in the CALYPSO code[48, 49]. The structural relaxations and electronic properties were calculated using density functional theory with the Perdew-Burke-Ernzerhof generalized gradient approximation as implemented in the VASP code[50,51]. The ion-electron interaction was described by projector-augmented-wave potentials, where $5s^25p^65d^16s^2$, $2s^22p^1$ and $1s^1$ configurations were treated as valence electrons for La, B and H atoms, respectively[52]. Plane wave kinetic energy cutoff was set to 700 eV and corresponding Monkhorst-Pack (MP) $k$-point meshes for different structures were adopted to ensure that the enthalpy converges to 1 meV/atom. Phonon calculations were performed by using the supercell method or density functional perturbation theory (DFPT) with PHONOPY[53] and Quantum-ESPRESSO codes[54], respectively. Electron-phonon coupling (EPC) calculations were carried out with the Quantum-ESPRESSO code using ultrasoft pseudopotentials for all atoms. We adopted a kinetic energy cutoff of 60 Ry. 7×7×5, and 9×9×9 $q$-point meshes in the first Brillouin zones (BZ) were used for $P\bar{3}m1$-LaBH$_7$ and $Fm\bar{3}m$-LaBH$_8$, respectively. Correspondingly, we chose MP grids of 28×28×20 and 36×36×36 to ensure $k$-point sampling convergence.

**Data availability**

The authors declare that the data supporting the findings of this study are available within the paper and its supplementary information files.

**References**

1. Ashcroft, N. W. Hydrogen Dominant Metallic Alloys: High Temperature Superconductors? *Phys. Rev. Lett.* **92**, 187002 (2004).
2. Wang, H., Li, X., Gao, G., Li, Y. & Ma, Y. Hydrogen-rich superconductors at high pressures. *WIREs: Comput. Mol. Sci.* **8**, e1330 (2018)
3. Zurek, E. & Bi, T. High-temperature superconductivity in alkaline and rare earth polyhydrides at high pressure: A theoretical perspective. *J Chem. Phys.* **150**, 050901 (2019).
4. Flores-Livas, J. A., Boeri, L., Sanna, A., Profeta, G., Arita, R.& Eremets, M. A perspective on conventional high-temperature superconductors at high pressure: Methods and materials, *Phys. Rep.* **856**, 1-78 (2020).
5. Semenok, D.V., Kruglov, I. A., Savkin, I. A., Kvashnin, A. G. & Oganov, A. R. On distribution of superconductivity in metal hydrides, *Curr. Opin. Solid State Mater. Sci.* **24**, 100808 (2020).
6. Pickard, C. J., Errea, I. & Eremets, M. I. Superconducting hydrides under pressure, *Annu. Rev. Conden. Matter Phys.* **11**, 57-76 (2020).
7. Duan, D. *et al.* Pressure-induced metallization of dense $(H_2S)_2H_2$ with high-$T_c$ superconductivity, *Sci. Rep.* **4**, 6968 (2014).
8. Wang, H., Tse, J. S., Tanaka, K., Iitaka, T. & Y. Ma, Superconductive sodalite-like clathrate calcium hydride at high pressures, *Proc. Natl. Acad. Sci. USA* **109**, 6463-6466 (2012).
9. Li, Y., Hao, J., Liu, H., Tse, J. S., Wang, Y. & Ma, Y. Pressure-stabilized superconductive yttrium hydrides, *Sci. Rep.* **5**, 9948 (2015).
10. Liang, X. et al. Potential high-$T_c$ superconductivity in $CaYH_{12}$ under pressure, *Phys. Rev. B* **99**, 100505 (2019).
11. Liu, H., Naumov, I. I., Hoffmann, R., Ashcroft, N. W. & Hemley, R.J. Potential high-Tc superconducting lanthanum and yttrium hydrides at high pressure, *Proc. Natl. Acad. Sci. USA* **114**, 6990-6995 (2017).
12. Peng, F., Sun, Y., Pickard, C. J., Needs, R. J., Wu, Q. & Ma,Y. Hydrogen clathrate structures in rare earth hydrides at high pressures: Possible route to room-temperature superconductivity, *Phys. Rev. Lett.* **119**, 107001 (2017).
13. Drozdov, A. P., Eremets, M. I., Troyan, I. A., Ksenofontov, V., Shylin, S. I. Conventional superconductivity at 203 kelvin at high pressures in the sulfur hydride system, *Nature* **525**, 73 (2015).
14. Einaga, M. et al. Crystal structure of the superconducting phase of sulfur hydride, *Nat. Phys.* **12**, 835 (2016).
15. Geballe, Z. M. et al. Synthesis and stability of lanthanum superhydrides, *Angew. Chem. Int. Ed.* **57**, 688-692 (2018).
16. Somayazulu, M.et al. Evidence for superconductivity above 260 K in lanthanum superhydride at megabar pressures, *Phys. Rev. Lett.* **122**, 027001 (2019).
17. Drozdov, A. P. et al. Superconductivity at 250 K in lanthanum hydride under high pressures, *Nature* **569**, 528-531 (2019).
18. Kong, P. et al. Superconductivity up to 243 K in yttrium hydrides under high pressure arXiv:1909.10482.

19. Troyan, I. A. et al. Anomalous High-temperature superconductivity in YH$_6$, *Adv. Mater.* 2006832 (2021).
20. Semenok, D. V. et al. Superconductivity at 161 K in thorium hydride ThH$_{10}$: Synthesis and properties, *Mater. Today* **33**, 36-44 (2020).
21. Snider, E. et al. Room-temperature superconductivity in a carbonaceous sulfur hydride, *Nature* **586**, 373-377 (2020).
22. Hutcheon, M. J., Shipley, A. M. & Needs, R. J. Predicting novel superconducting hydrides using machine learning approaches, *Phys. Rev. B* **101**, 144505 (2020).
23. Kruglov, I. A. et al. Uranium polyhydrides at moderate pressures: Prediction, synthesis, and expected superconductivity, *Sci. Adv.* **4,** eaat9776 (2018).
24. Guigue, B., Marizy, A. & Loubeyre, P. Synthesis of UH$_7$ and UH$_8$ superhydrides: Additive-volume alloys of uranium and atomic metal hydrogen down to 35 GPa, *Phys. Rev. B* **102**, 014107 (2020).
25. Salke, N. P. et al. Synthesis of clathrate cerium superhydride CeH$_9$ at 80-100 GPa with atomic hydrogen sublattice. *Nature Commun.* **10**, 4453 (2020).
26. Chen, W. et al. Synthesis of molecular metallic barium superhydride: pseudocubic BaH$_{12}$, *Nature Commun.* **12**, 1-9 (2021).
27. Cui, W. et al. Route to high-$T_c$ superconductivity via CH$_4$-intercalated H$_3$S hydride perovskites, *Phys. Rev. B* **101**, 134504 (2020).
28. Sun, Y. et al. Computational discovery of a dynamically stable cubic SH$_3$-like high-temperature superconductor at 100 GPa via CH$_4$ intercalation, *Phys. Rev. B* **101**, 174102 (2020).
29. Ge, Y., Zhang, F., Dias, R. P., Hemley, R. J. & Yao, Y. Hole-doped room-temperature superconductivity in H$_3$S$_{1-x}$Z$_x$ (Z=C, Si). *Mater. Today Phys.* **15**, 100330 (2020).
30. Kokail, C., von der Linden, W. & Boeri, L. Prediction of high-$T_c$ conventional superconductivity in the ternary lithium borohydride system, *Phys. Rev. Mater.* **1**, 074803 (2017).
31. Di Cataldo, S., von der Linden, W. & Boeri, L. Phase diagram and superconductivity of calcium borohyrides at extreme pressures, *Phys. Rev. B* **102**, 014516 (2020).
32. Hu, C.-H. et al. Pressure-induced stabilization and insulator-superconductor transition of BH, *Phys. Rev. Lett.* **110**, 165504 (2013).
33. Yao, Y. & Hoffmann, R. BH$_3$ under pressure: Leaving the molecular diborane motif, *J. Am. Chem. Soc.* **133**, 21002-21009 (2011).
34. Yang, W.-H. et al. Novel superconducting structures of BH$_2$ under high pressure, *Phys. Chem. Chem. Phys.* **21**, 5466-5473 (2019).
35. Grockowiak, A. et al. Hot hydride superconductivity above 550 K, arXiv:2006.03004.
36. Liu, H., Naumov, I. I., Geballe, Z. M., Somayazulu, M., Tse, J. S. & Hemley, R. J. Dynamics and superconductivity in compressed lanthanum superhydride. *Phys. Rev. B* **98**, 100102 (2018).
37. Errea, I. et al. Quantum crystal structure in the 250-kelvin superconducting lanthanum hydride. *Nature* **578,** 66-72 (2019).

38. Kruglov, I. A. et al. Superconductivity of LaH$_{10}$ and LaH$_{16}$ polyhydrides, *Phys. Rev. B* **101**, 024508 (2020).
39. Ma, Y. & Tse, J. S. Ab initio determination of crystal lattice constants and thermal expansion for germanium isotopes, *Solid State Commun.* **143**, 161-165 (2007).
40. Becke, A. D. & Edgecombe, K. E. A simple measure of electron localization in atomic and molecular systems, *J Chem. Phys.* **92**, 5397-5403 (1990).
41. Savin, A., Jepsen, O., Flad, J., Andersen, O.K., Preuss, H. & von Schnering, H.G. Electron Localization in Solid-State Structures of the Elements: the Diamond Structure, *Angew. Chem. Int. Ed.* **31**, 187-188 (1992).
42. Bader, R. Atoms in Molecules: A Quantum Theory (Oxford University Press, Oxford, UK, 1994).
43. Simon, A. Superconductivity and chemistry, *Angew. Chem. Int. Ed.* **36**, 1788-1806 (1997).
44. Allen, P. B. & Dynes, R. C. Transition temperature of strong-coupled superconductors reanalyzed, *Phys. Rev. B* **12**, 905-922 (1975).
45. Eliashberg, G. M. Interactions between electrons and lattice vibrations in a superconductor, *Sov. Phys. JETP* **11,** 696-702 (1960).
46. Di Cataldo, S., Heil, G., von der Linden, W. & Boeri, L. LaBH$_8$: the first high-$T_c$ low-pressure superhydride, arXiv:2102.11227.
47. Wang, H., Yao, Y., Peng, F., Liu, H. & R. J. Hemley, Quantum and classical proton diffusion in superconducting clathrate hydrides, *Phys. Rev. Lett.* **126**, 117002 (2021).
48. Wang, Y., Lv, J., Zhu, L. & Ma, Y. Crystal structure prediction via particle-swarm optimization, *Phys. Rev. B* **82**, 094116 (2010).
49. Wang, Y., Lv, J., Zhu, L. & Ma, Y. CALYPSO: A method for crystal structure prediction, *Comput. Phys. Commun.* **183**, 2063-2070 (2012).
50. Kresse, G. & Furthmüller, J. Efficient iterative schemes for ab initio total-energy calculations using a plane-wave basis set, *Phys. Rev. B* **54**, 11169-11186 (1996).
51. Perdew, J. P., Chevary, J. A., Vosko, S. H., Jackson, K. A., Pederson, M. R., Singh, D. J. & Fiolhais, C. Atoms, molecules, solids, and surfaces: Applications of the generalized gradient approximation for exchange and correlation, *Phys. Rev. B* **46**, 6671-6687 (1992).
52. Blöchl, P. E. Projector augmented-wave method, *Phys. Rev. B* **50**, 17953-17979 (1994).
53. Togo, A., Oba, F. & Tanaka, I. First-principles calculations of the ferroelastic transition between rutile-type and CaCl$_2$-type SiO$_2$ at high pressures, *Phys. Rev. B* **78**, 134106 (2008).
54. Paolo, G. et al. QUANTUM ESPRESSO: a modular and open-source software project for quantum simulations of materials, *J. Phys.: Condens. Matter* **21**, 395502 (2009).

**Acknowledgements**

The work was supported by National Natural Science Foundation of China (No. 52022089, 12074138), and the Ph.D. Foundation by Yanshan University (Grant No.

B970). A.B. acknowledges financial support from the Spanish Ministry of Science and Innovation (Grant No. FIS2019-105488GB-I00). R.H. acknowledges support from the U.S. National Science Foundation (DMR-1933622).

**Author contributions**

G.G. conceived this project. X.L, X.W., L.W. and R.S. performed the calculations and analysis. X.L., A.B., H.L., R.J.H., L.W., G.G. and Y.T. wrote and revised the paper. All the authors discussed the results and offered useful inputs.

**Competing interests**

The authors declare no competing interests.

**Additional information**

Supplementary information is available for this paper at ***.

**Correspondence and requests for materials should be addressed to G.G. or H.L.**